\documentstyle[editedvolume,psfig]{crckapb10}
\begin{opening}
\title{1.4-GHz LUMINOSITY FUNCTION OF GALAXIES\protect\\
      FROM THE LAS CAMPANAS REDSHIFT SURVEY}
\author{J. MACHALSKI, W. GODLOWSKI}
\institute{Astronomical Observatory, Jagellonian University\\
          ul. Orla 171, PL-30244 Cracow, Poland}
\end{opening}

\runningtitle{1.4-GHz LUMINOSITY FUNCTION}

\begin{document}

\begin{abstract}
A preliminary 1.4 GHz RLF at redshift of about 0.14 is derived from the 
{\it Las Campanas Redshift Survey} (LCRS) and the NVSS radio data. No 
significant evolution has been found at this redshift in comparison to the 
'local' RLF.
\end{abstract}

The LCRS consists of over 26000 redshifts of galaxies with 
$15.0< R(mag)< 17.7$ lying in six sky strips between declinations 
$(-3^{o},-45^{o})$ (Shectman et al. 1996, ApJ,470,172). 2/3 of the strips 
are covered by the NVSS survey. The radio data released up to date have 
allowed only 15 per cent of the optical survey to be used for the RLF 
determination.

The optical positions of 11671 LCRS galaxies were searched to a distance of 
15 arc sec in the NVSS source list. 95 of about 7000 galaxies with 
$R\leq 17.7$ mag were detected above the NVSS flux limit of 2.5 mJy. Additional 
33 detections were found between $17.7< R(mag)< 18.5$. Originally LBDS 
galaxies that met the given photometric criteria had been chosen at random for 
the multi-object spectroscopy. Therefore about 2/3 of the galaxies identified 
with radio sources are not in the redshift catalogue. In order to use them 
in construction of the RLF, their photometric redshift was estimated. The 
resultant distribution of z is shown in Fig.1(a).

The {\it far infrared} (FIR) identification of galaxies allows to select 'starbursts' 
from among all galaxies. Using the IRAS Faint and Point Source Catalogues, we 
were able to identify 30 of 128 radio-detected LCRS galaxies. All of them but one 
have $q> 1.9$ $[q\equiv log(L_{FIR}/L_{20cm}]$ (Fig.1b), 
i.e. they can be assumed to be 'pure' starburst galaxies. Other 22 galaxies likely 
have the same type. Oppositely, LCRS galaxies with an upper limit $q< 1.1$, 
as well as these with the degree of polarized 1.4 GHz flux exceeding 10 per 
cent of the total, can be considered as the 'AGN' type, powered mostly by 
nuclear processes.

\begin{figure}
\vspace*{-3.5cm}
\centering{\psfig{figure=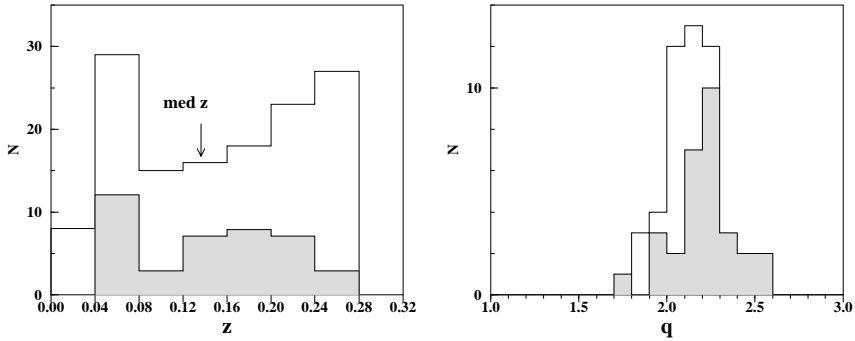,width=11.0cm}}
\vspace{-0.5cm}
\caption{\label{fig} (a) n(z) of radio-identified LCRS galaxies. Spectroscopic 
redshifts are hatched; (b) n(q) of {\it IRAS}-identified galaxies. All of them 
have $q>1.9$, thus they have to be 'starbursts'. The unhatched area contains 
possible 'starbursts' with an upper limit of q}
\end{figure}

\underline {Preliminary} RLFs were determined for all LCRS galaxies, as well as for 
'starburst' and 'AGN' subdivision using the method of Condon (1989, ApJ,338,13). 
The resultant functions are shown in Figs.2(a),2(b).

\begin{figure}
\centerline{
\psfig{figure=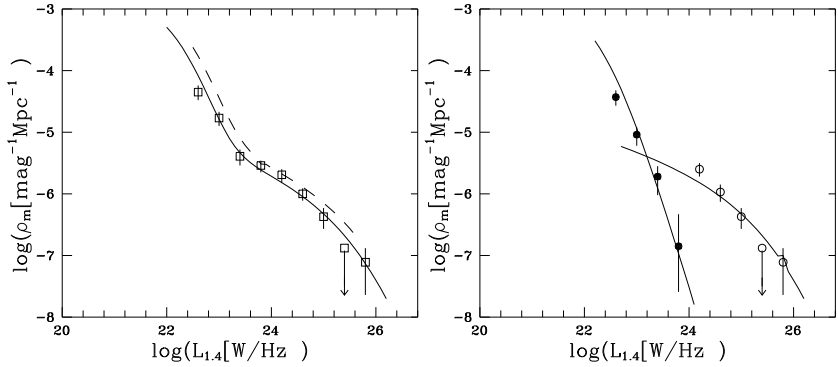,width=11.0cm}
}
\caption{\label{fig1} (a) The 1.4 GHz RLF at $z\approx 0.14$ determined from 95 LCRS 
galaxies with $R\leq 17.7$. Solid curve shows the 'local' RLF $(z\approx 0.02)$ 
determined by Condon (1989); dashed curve shows its evolution to $z\approx 0.14$ 
predicted from the C84 model; (b) The RLF of 'starburst' (filled dots) and 
'AGN' (open circles) LCRS galaxies. The solid curves show the corresponding 
'local' RLFs determined by Condon (1989)}
\end{figure}

The volume of space sampled is radio limited for the radio luminosities below 
$L\approx 10^{23.6}$ W/Hz. Above $10^{23.6}$ W/Hz the limiting volume is always 
determined optically and reaches redshifts $z\approx 0.26$. Consequently the 
RLF is well determined between luminosities $10^{22.4}$ to $10^{25}$ W/Hz. 
The resultant preliminary RLFs appear undistinguishable from the 
'local' ones at $z\approx 0.02$ (Condon 1989). Presently available NVSS data 
are insufficient to check whether any evolution of galaxies at $z\approx 0.14$ 
has taken place, and if so, whether its amount  is consistent with the model 
predictions. This, however, should be possible when the remaining 85 per cent 
of the NVSS data become available.

\end{document}